# Analytical Solutions of Classical and Fractional KP-Burger Equation and Coupled KdV equation

Uttam Ghosh[1a], Susmita Sarkar[1b] and Shantanu Das[3]

[1]Department of Applied Mathematics, University of Calcutta, Kolkata, India
[1a]email : uttam_math@yahoo.co.in
[1b]email : susmita62@yahoo.co.in
[3]Reactor Control Systems Design Section E & I Group BARC Mumbai India
email : shantanu@barc.gov.in

**Abstract**

Evaluation of analytical solutions of non-linear partial differential equations (both classical and fractional) is a rising subject in Applied Mathematics because its applications in Physical biological and social sciences. In this paper we have used generalized Tanh method to find the exact solution of KP-Burger equation and coupled KdV equation. The fractional Sub-equation method has been used to find the solution of fractional KP-Burger equation and fractional coupled KdV equations. The exact solution obtained by fractional sub-equation method reduces to classical solution when order of fractional derivative tends to one. Finally numerical simulation has done. The numerical simulation justifies that the solutions of two fractional differential equations reduces to shock solution for KP-Burger equation and soliton solution for coupled KdV equations when order of derivative tends to one.

**Keywords:** Generalized Tanh-method, Fractional Sub-Equation Method, KP-Burger equation, Coupled KdV equation, Fractional Differential Equation, Jumarie Fractional Derivative.

**1.0 Introduction**

Exact solutions of non-linear differential equations give complete picture of physical systems which cannot be obtained from their linear approximation. But it is very difficult to find the exact solutions of non-linear differential equations. There are many approximate methods to find the solutions of the non-linear differential equations. The approximate methods are Adomain Decomposition Method[1-4], Homotopy Perturbation Method (HPM)[5-7], Differential Transform Method (DTM)[8] etc. Currently researcher in this field is developing new methods to find the exact solutions of non-linear differential equations. The Tanh method was introduced by Huiblin and Kelin[9] to find the travelling wave solutions of non-linear differential equations. Wazwaz [10] used this method to find soliton solutions of the Fisher equation in analytic form. Fan [11] modified this Tanh method to solve KdV-Burgers-Kuamoto equations and Boussinesq equation.

Another growing field of applied science and engineering is the fractional calculus [12] where physical processes are studied in terms of the fractional differential equations. Zhang and Zhang [13] developed the fractional sub-equation method to find the travelling wave solutions of the Jumarie type fractional differential [14] equation in terms of the fractional tanh functions. The fractional sub-equation method and Generalized Tanh –method both are based on the Homogeneous balance principal [9] . The fractional sub-equations methods are used by authors to solve different non-linear fractional differential equations. Recently we have developed an algorithm to solve the linear fractional differential equations in terms of one parameter Mittag-Leffler function[15]. In this paper we shall use Generalized Tanh method and Fractional Sub-equation method to exact solutions of KP-Burger and coupled KdV equations and the corresponding fractional differential equation. Using these methods we obtain the soliton solution and periodic solutions. Organization of the paper is as follows. In section 2.0 we describe the principle of Tanh method and fractional sub-equation method. In section 3.0 we found the solutions of the KP-Burgers equation, in section 4.0 we found the solutions of the fractional order KP-Burgers equations. In section 5.0 we found the solutions of the coupled KdV equations, in section 6.0 we found the solutions of the fractional order coupled KdV equations. Finally numerical simulations are done for different values of the fractional order derivative.

**2.0 Generalized Tanh Method and fractional Sub-equation Method**

**(a) Generalized Tanh method**

In this method the solutions of the non-linear partial differential equations are expressed in terms of tanh and tan-functions. Consider the non-linear partial differential equation

$$L(u, u_t, u_x, u_y, u_{tt}, u_{xx}, u_{yy}...) = 0 \qquad (1)$$

satisfied by $u(x, y, t)$. Use of the travelling wave transformation in the form $\xi = kx + my + ct$, where $(k, m)$ are the wave vector and $c$ is the velocity of propagating waves, equation (1) reduces to

$$L(u, u', u'', ...) = 0 \qquad (2)$$

This is an ordinary differential equation of $u(\xi)$. The generalized tanh method of Fan and Hon[16] is based on the priori assumption that the travelling wave solutions can be expressed as the power series expansion which is the solution of non-linear Riccati differential equation $\phi'(\xi) = \sigma + \phi^2$. Solution of this equation can be written in the form

$$\phi(\xi) = \begin{cases} \begin{rcases} -\sqrt{-\sigma}\tanh(\sqrt{-\sigma}\xi) \\ -\sqrt{-\sigma}\coth(\sqrt{-\sigma}\xi) \end{rcases} & \text{for } \sigma < 0 \\ \begin{rcases} \sqrt{\sigma}\tan(\sqrt{\sigma}\xi) \\ -\sqrt{\sigma}\cot(\sqrt{\sigma}\xi) \end{rcases} & \text{for } \sigma > 0. \\ -\dfrac{1}{\xi} & \text{for } \sigma = 0 \end{cases} \qquad (3)$$

Let $u = S(\phi) = a_0 + a_1\phi + a_2\phi^2 + ... + a_n\phi^n$ be the solution of the equation (2) where $\phi(\xi)$ is given by (3) and $a_0, a_1, ....$ are constants. Then $u' = (a_1 + 2a_2\phi + ... + na_n\phi^{n-1})(\sigma + \phi^2)$ has highest power of $\phi$ as $n+1$. Similarly the $u''$ has highest power of $\phi$ as $n+2$. Then equating the highest power of $\phi$ from the highest order derivative term and the non-linear term the value of $n$ can be obtained. Then putting $S(\phi)$ in equation (2) and equating the like powers of $\phi$ the values of the values of $a_0, a_1, ....$ can be determine.

**(b) Fractional sub-equation methods**

The non-linear fractional partial differential equation is of the form,

$$L(u, u_t^{(\alpha)}, u_x^{(\alpha)}, u_{xx}^{(2\alpha)}, u_y^{(\alpha)}, u_{yy}^{(2\alpha)}.....) = 0 \qquad 0 < \alpha \leq 1 \qquad (4)$$

where $u = u(x, y, t)$ and $L$ is linear or non-linear operator. $\alpha$ is the order of the fractional derivative of Jumarie type defined as follows

$$_0^J D_x^\alpha [f(x)] = f^{(\alpha)}(x) = \begin{cases} \dfrac{1}{\Gamma(-\alpha)} \int_0^x (x-\xi)^{-\alpha-1} f(\xi) d\xi, & \alpha < 0 \\ \dfrac{1}{\Gamma(1-\alpha)} \dfrac{d}{dx} \int_0^x (x-\xi)^{-\alpha} (f(\xi) - f(0)) d\xi, & 0 < \alpha < 1 \\ \left(f^{(\alpha-n)}(x)\right)^{(n)}, & n \leq \alpha < n+1, \quad n \geq 1 \end{cases}$$

We point out that composition as inequality $D^\alpha D^\alpha \neq D^{2\alpha}$, holds.

We consider that at $x < 0$ the function $f(x) = 0$ and also $f(x) - f(0) = 0$ for $x < 0$. The fractional derivative considered here in the fractional differential equation are obtained using Jumarie [6] modified Riemann-Liouville (RL) derivative as defined above. The first expression above is fractional integration of Jumarie type. The modification by Jumarie is to carry RL fractional integration or RL fractional differentiation by forming a new function, offsetting the original function by subtraction of the function value at the start point; and then operate the RL definition. Using the Jumarie type derivative the following can be obtained [14]

$$_0^J D_x^\alpha [x^\gamma] = \frac{\Gamma(1+\gamma)}{\Gamma(1+\gamma-\alpha)} x^{\gamma-\alpha}, \qquad \gamma > 0,$$

$$_0^J D_x^\alpha [f(x)g(x)] = g(x)\left(_0^J D_x^\alpha f(x)\right) + f(x)\left(_0^J D_x^\alpha g(x)\right)$$

$$_0^J D_x^\alpha [f(g(x))] = f_g'(g(x))\left(_0^J D_x^\alpha [g(x)]\right) = \left(_0^J D_g^\alpha [f(g(x))]\right)(g_x')^\alpha$$

Then using the travelling wave transformation $\xi = kx + my + ct$ equation (4) reduces to

$$L(u, u_\xi^{(\alpha)}, u_{\xi\xi}^{(2\alpha)}.....) = 0 \qquad 0 < \alpha \leq 1 \qquad (5)$$

Whose solution can be expressed in the form $u = S(\varphi) = a_0 + a_1\varphi + a_2\varphi^2 + ... + a_n\varphi^n$ where $\varphi$ satisfies the fractional Riccati equation $D^\alpha \phi(\xi) = \sigma + \phi^2$, $0 < \alpha \leq 1$ and $a_i$'s are arbitrary constants, Zhang et al [8] established generalized exp-method solution of the fractional differential equation $D^\alpha \phi(\xi) = \sigma + \phi^2$, $0 < \alpha \leq 1$ in the form,

$$\phi(\xi) = \begin{cases} -\sqrt{-\sigma}\tanh_\alpha(\sqrt{-\sigma}\xi) \\ -\sqrt{-\sigma}\coth_\alpha(\sqrt{-\sigma}\xi) \end{cases} \quad \text{for } \sigma < 0 \\ \begin{cases} \sqrt{\sigma}\tan_\alpha(\sqrt{\sigma}\xi) \\ -\sqrt{\sigma}\cot_\alpha(\sqrt{\sigma}\xi) \end{cases} \quad \text{for } \sigma > 0 \\ -\dfrac{\Gamma(1+\alpha)}{\xi^\alpha + \omega} \quad \text{for } \sigma = 0, \omega = \text{Constant} \end{cases} \qquad (6)$$

where the fractional trigonometric functions and fractional hyperbolic functions are defined in [8] in the form,

$$\tanh_\alpha(x) = \frac{\sinh_\alpha(x)}{\cosh_\alpha(x)} \qquad \coth_\alpha(x) = \frac{\cosh_\alpha(x)}{\sinh_\alpha(x)}$$

$$\sinh_\alpha(x) = \frac{E_\alpha(x^\alpha) - E_\alpha(x^\alpha)}{2} \qquad \cosh_\alpha(x) = \frac{E_\alpha(x^\alpha) + E_\alpha(x^\alpha)}{2}$$

$$\tan_\alpha(x) = \frac{\sin_\alpha(x)}{\cos_\alpha(x)} \qquad \cot_\alpha(x) = \frac{\cos_\alpha(x)}{\sin_\alpha(x)}$$

$$\sin_\alpha(x) = \frac{E_\alpha(ix^\alpha) - E_\alpha(ix^\alpha)}{2i} \qquad \cos_\alpha(x) = \frac{E_\alpha(ix^\alpha) + E_\alpha(ix^\alpha)}{2}$$

Where $E_\alpha(z) = \sum_{k=0}^{\infty} \frac{z^\alpha}{\Gamma(1+k\alpha)}$ is the one parameter Mittag-Leffler function.

Using the above described methods we find the analytic solutions of the non-linear (I) KP-Burger equations in 2+1 dimensions and (II) Coupled KdV equations and the corresponding space and time fractional differential equations.

### 3.0 Generalized Solutions of the KP-Burgers equation using Generalized Tanh method.

Let us consider the 2+1 dimensional KP-Burger equation satisfied $u = u(x, y, t)$ in of the form

$$(u_t + uu_x + pu_{xxx} - qu_{xx})_x + ru_{yy} = 0 \qquad (7)$$

Using the travelling wave transformation $\xi = lx + my + ct$ equation (7) reduces to the non-linear ordinary differential equation,

$$l(cu_\xi + luu_\xi + pl^3 u_{\xi\xi\xi} - l^2 qu_{\xi\xi})_\xi + rm^2 u_{\xi\xi} = 0 \qquad (8)$$

where $l, m, c$ are constants.

Now using the localized boundary condition $u(\xi) \to 0$ for $\xi \to \pm\infty$, integration of the equation (8) w.r.to $\xi$ gives,

$$l\left(cu + l\frac{u^2}{2} + pl^3 u_{\xi\xi} - l^2 q u_\xi\right) + rm^2 u = 0$$

$$\text{or, } 2(lc + rm^2)u + lu^2 - 2l^2 q u_\xi + 2pl^3 u_{\xi\xi} = 0 \quad (10)$$

Which is a non-linear ordinary differential equation satisfied by $u(\xi)$. Now we solve the above equation using the generalized Tanh method.

For this purpose let us consider $u(\xi) = S(\phi(\xi)) = a_0 + a_1\phi + a_2\phi^2 + \ldots + a_n\phi^n$ be a series solution of the differential equation (10) where $\phi(\xi)$ satisfies the Riccati differential equation $\phi'(\xi) = \sigma + \phi^2$ and $a_i$'s arbitrary constants [10]. Putting this in equation (10) and using the principle of homogeneous balance we compare the highest power of $\phi(\xi)$ from the non-linear term and the highest order derivative term of $\phi(\xi)$. We thus get $n = 2$.

Therefore the series solution (10) reduces to $u(\xi) = a_0 + a_1\phi + a_2\phi^2$ with $a_2 \neq 0$. Now putting this in equation (10) we get

$$2(lc + rm^2)(a_0 + a_1\varphi + a_2\varphi^2) + l(a_0 + a_1\varphi + a_2\varphi^2)^2 - 2l^2 q(a_1 + 2a_2\varphi)(\sigma + \phi^2)$$
$$+ 2pl^3(2a_2\sigma^2 + 2a_1\sigma\varphi + 8a_2\sigma\varphi^2 + 2a_1\varphi^3 + 6a_2\varphi^4) = 0 \quad (11)$$

Comparing the like powers of $\phi$ we get

$$\phi^0 : 2(lc + rm^2)a_0 + la_0^2 - 2ql^2 a_1 \sigma + 4pa_2 l^3 = 0$$
$$\phi^1 : 2(lc + rm^2)a_1 + 2la_0 a_1 - 4a_2 ql^2 \sigma + 4a_1 pl^3 \sigma = 0$$
$$\phi^2 : 2(lc + rm^2)a_2 + la_1^2 + 2la_0 a_2 - 2a_1 ql^2 + 16a_2 pl^3 \sigma = 0\ldots \quad (12)$$
$$\phi^3 : 2la_1 a_2 - 4ql^2 a_2 + 4pl^3 a_1 = 0$$
$$\phi^4 : la_2^2 + 12a_2 pl^3 = 0$$

Solving the above we get $a_2 = -12pl^2$, $a_1 = \frac{12}{5}ql$ and $a_0 = -8pl^2\sigma - \frac{(cl + rm^2)}{l} + \frac{q^2}{25p}$

The general solution of the above equations is

$$u(x,y,t) = \begin{cases} a_0 - a_1\sqrt{-\sigma}\tanh(\sqrt{-\sigma}(ct+lx+my)) + 12pl^2\sigma\tanh^2\left(\sqrt{-\sigma}(ct+lx+my)\right) \\ a_0 - a_1\sqrt{-\sigma}\coth(\sqrt{-\sigma}(ct+lx+my)) + 12pl^2\sigma\coth^2\left(\sqrt{-\sigma}(ct+lx+my)\right) \end{cases} \text{ for } \sigma < 0 \\ a_0 - a_1\frac{1}{(ct+lx+my)} - 12pl^2\left(\frac{1}{((ct+lx+my))^2}\right) \quad \text{ for } \sigma = 0 \\ \begin{cases} a_0 + a_1\sqrt{\sigma}\tan(\sqrt{\sigma}(ct+lx+my)) - 12pl^2\sigma\tan^2\left(\sqrt{\sigma}(ct+lx+my)\right) \\ a_0 - a_1\sqrt{\sigma}\cot(\sqrt{\sigma}(ct+lx+my)) - 12pl^2\sigma\coth^2\left(\sqrt{\sigma}(ct+lx+my)\right) \end{cases} \text{ for } \sigma > 0 \end{cases} \quad \ldots(13)$$

This is generalized solution of the KP-Burger equation in 2+1 dimension. The first two solutions are sock solutions and the last two solutions are the periodic solutions.

**4.0 Solutions of the fractional order KP-Burgers equation using fractional sub equation method**

The fractional KP-Burger equation in 2+1 dimension is of the form

$$\left(u_t^{(\alpha)} + uu_x + pu_{xxx}^{(3\alpha)} - qu_{xx}^{(2\alpha)}\right)_x^{(\alpha)} + ru_{yy}^{(2\alpha)} = 0 \tag{14}$$

where $u_{yy}^{(2\alpha)} = \frac{\partial^{2\alpha} u}{\partial y^{2\alpha}}$, $u_{xxx}^{(3\alpha)} = \frac{\partial^{3\alpha} u}{\partial t^{3\alpha}}$, $u_t^{(\alpha)} = \frac{\partial^{\alpha} u}{\partial t^{\alpha}}$, $0 < \alpha \leq 1$

Using the travelling wave transformation $\xi = lx + my + ct$ equation (14) reduces to

$$l^{\alpha}\left(c^{\alpha}u_\xi^{(\alpha)} + l^{\alpha}uu_\xi^{(\alpha)} + pl^{3\alpha}u_{\xi\xi\xi}^{(3\alpha)} - l^{2\alpha}qu_{\xi\xi}^{(2\alpha)}\right)_\xi + rm^{2\alpha}u_{\xi\xi}^{(2\alpha)} = 0 \tag{15}$$

where $l, m, c$ are constants.

Integrating fractionally twice both sides of (15) with respect to $\xi$ and using the localized conditions of solitary waves i.e. $u_\xi^{(\alpha)}$ and $u(\xi) \to 0$ for $\xi \to \pm\infty$ we get,

$$2(l^{\alpha}c^{\alpha} + rm^{2\alpha})u + l^{\alpha}u^2 - 2l^{2\alpha}qu_\xi^{(\alpha)} + 2pl^{3\alpha}u_{\xi\xi}^{(2\alpha)} = 0 \tag{16}$$

This is a non-linear ordinary fractional differential equation. The above equation will be solved using the fractional sub-equation method,

Consider solution of the equation (16) in the form $u = S(\varphi(\xi)) = a_0 + a_1\varphi + a_2\varphi^2 + \ldots + a_n\varphi^n$ where $\varphi(\xi)$ satisfies the fractional Riccati equation $D^{\alpha}\phi(\xi) = \sigma + \phi^2, 0 < \alpha \leq 1$ [13] and $a_i$'s are arbitrary constants. Using the homogeneous balance principle we get $n = 2$.

Thus we have solutions in the form $u = a_0 + a_1\varphi + a_2\varphi^2$, putting this in equation (16) we get,

$$2(l^\alpha c^\alpha + rm^{2\alpha})(a_0 + a_1\varphi + a_2\varphi^2) + l^\alpha(a_0 + a_1\varphi + a_2\varphi^2)^2 - 2l^{2\alpha}q(a_1 + 2a_2\varphi)(\sigma + \phi^2)$$
$$+ 2pl^{3\alpha}(2a_2\sigma^2 + 2a_1\sigma\varphi + 8a_2\sigma\varphi^2 + 2a_1\varphi^3 + 6a_2\varphi^4) = 0 \tag{17}$$

Comparing the like powers of $\varphi$ from both sides we get

$$\varphi^0 : 2(l^\alpha c^\alpha + rm^{2\alpha})a_0 + l^\alpha a_0^2 - 2ql^{2\alpha}a_1\sigma + 4pa_2l^{3\alpha}\sigma^2 = 0$$
$$\varphi^1 : 2(l^\alpha c^\alpha + rm^{2\alpha})a_1 + 2l^\alpha a_0 a_1 - 4a_2 q l^{2\alpha}\sigma + 4a_1 p l^{3\alpha}\sigma = 0$$
$$\varphi^2 : 2(l^\alpha c^\alpha + rm^{2\alpha})a_2 + l^\alpha a_1^2 + 2l^\alpha a_0 a_2 - 2a_1 q l^{2\alpha} + 16 a_2 p l^{3\alpha}\sigma = 0 \ldots \tag{18}$$
$$\varphi^3 : 2l^\alpha a_1 a_2 - 4ql^{2\alpha}a_2 + 4pl^{3\alpha}a_1 = 0$$
$$\varphi^4 : l^\alpha a_2^2 + 12 a_2 p l^{3\alpha} = 0$$

Solution of (18) for $a_0, a_1, a_2$ give

$$a_2 = -12pl^{2\alpha}, \quad a_1 = \frac{12}{5}ql^\alpha \text{ and } a_0 = -8\sigma p l^{2\alpha} - \frac{(c^\alpha l^\alpha + rm^{2\alpha})}{l^\alpha} + \frac{q^2}{25p} \ldots \tag{19}$$

Thus the general solution of the above equations is

$$u(x,y,t) = \begin{cases} \begin{aligned} &a_0 - a_1\sqrt{-\sigma}\tanh_\alpha(\sqrt{-\sigma}(c^\alpha t + k^\alpha x + m^\alpha y)) + 12pl^{2\alpha}\sigma\tanh_\alpha^2(\sqrt{-\sigma}(c^\alpha t + k^\alpha x + m^\alpha y)) \\ &a_0 - a_1\sqrt{-\sigma}\coth_\alpha(\sqrt{-\sigma}(c^\alpha t + k^\alpha x + m^\alpha y)) + 12pl^{2\alpha}\sigma\coth_\alpha^2(\sqrt{-\sigma}(c^\alpha t + k^\alpha x + m^\alpha y)) \end{aligned} & \text{for} \quad \sigma < 0 \\[2ex] a_0 - a_1\frac{(\Gamma(1+\alpha))}{\left((c^\alpha t + k^\alpha x + m^\alpha y)^\alpha + \omega\right)} - 12pl^{2\alpha}\frac{(\Gamma(1+\alpha))^2}{\left((c^\alpha t + k^\alpha x + m^\alpha y)^\alpha + \omega\right)^2} & \text{for} \quad \sigma = 0, \; \omega - \text{Constant} \quad \ldots (20) \\[2ex] \begin{aligned} &a_0 + a_1\sqrt{\sigma}\tan_\alpha(\sqrt{\sigma}(c^\alpha t + k^\alpha x + m^\alpha y)) - 12pl^{2\alpha}\sigma\tan_\alpha^2(\sqrt{\sigma}(c^\alpha t + k^\alpha x + m^\alpha y)) \\ &a_0 - a_1\sqrt{\sigma}\cot_\alpha(\sqrt{\sigma}(c^\alpha t + k^\alpha x + m^\alpha y)) - 12pl^{2\alpha}\sigma\coth_\alpha^2(\sqrt{\sigma}(c^\alpha t + k^\alpha x + m^\alpha y)) \end{aligned} & \text{for} \quad \sigma > 0 \end{cases}$$

This is the exact analytic solution of the fractional KP-Burger equation in 2+1 dimension.

**5.0 Generalized Solutions of the Coupled KdV equations using Generalized Tanh method**

Consider the coupled KdV equation with constant coefficients in the form

$$\left.\begin{aligned} u_t + avu_x + bu_{xxx} &= 0 \\ v_t + duv_x + bv_{xxx} &= 0 \end{aligned}\right\} \tag{21}$$

where $a, b, d$ are constants, they may be function of $t$ in some cases and $u = u(x,t), v = v(x,t)$. To find the solition solutions of the coupled differential equations (21) here Tanh method is used. The travelling wave transformation in the form $\xi = kx + ct$ where the constant k is called the wave number and another constant c is the velocity of the propagating wave the equation (21) reduces to

$$cu_\xi + akvu_\xi + bk^3 u_{\xi\xi\xi} = 0 \Big\}$$
$$cv_\xi + dkuv_\xi + bk^3 v_{\xi\xi\xi} = 0 \Big\} \qquad (22)$$

which are coupled non-linear ordinary differential equations with $u_\xi = \dfrac{du}{d\xi}$ and $v_\xi = \dfrac{dv}{d\xi}$. We want to find the series solution of the system of differential equation in the following form where $\varphi(\xi)$ satisfies the Riccati equation $D\phi(\xi) = \sigma + \phi^2$, and $a_i$ and $b_i$'s are arbitrary constants [10].

$$u(\xi) = S_1(\varphi) = a_0 + a_1\phi + a_2\phi^2 + \ldots + a_n\phi^n$$
$$v(\xi) = S_2(\varphi) = b_0 + b_1\phi + b_2\phi^2 + \ldots + b_n\phi^n$$

Using the homogeneous balance principle as previous we get $m = n = 2$. Thus we get

$$u(\xi) = S_1(\varphi) = a_0 + a_1\varphi + a_2\varphi^2 \Big\}$$
$$v(\xi) = S_2(\varphi) = b_0 + b_1\varphi + b_2\varphi^2 \Big\} \qquad (23)$$

Putting the above in equation (22) we get,

$$c(a_1\sigma + 2a_2\sigma\phi + a_1\phi^2 + 2a_2\phi^3) + ak(b_0 + b_1\phi + b_2\phi^2)(a_1\sigma + 2a_2\sigma\phi + a_1\phi^2 + 2a_2\phi^3)$$
$$+ bk^3(2a_1\sigma^2 + 16a_2\phi\sigma^2 + 8a_1\sigma\phi^2 + 40a_2\sigma\phi^3 + 6a_1\phi^4 + 24a_2\phi^5) = 0$$

and $c(b_1\sigma + 2b_2\sigma\phi + b_1\phi^2 + 2b_2\phi^3) + ak(b_0 + b_1\phi + b_2\phi^2)(b_1\sigma + 2b_2\sigma\phi + b_1\phi^2 + 2b_2\phi^3)$
$$+ bk^3(2b_1\sigma^2 + 16b_2\phi\sigma^2 + 8b_1\sigma\phi^2 + 40b_2\sigma\phi^3 + 6b_1\phi^4 + 24b_2\phi^5) = 0 \qquad (24)$$

Comparing the like powers of $\phi$ we get

For the first equation

$$\phi^0 : ca_1\sigma + akb_0 a_1\sigma + 2a_1 k^3 b\sigma^2 = 0$$
$$\phi^1 : 2ca_2\sigma + ak(2b_0 a_2 + b_1 a_1)\sigma + 16a_2 bk^3\sigma^2 = 0$$
$$\phi^2 : ca_1 + ak(b_0 a_1 + 2b_1 a_2\sigma + b_2 a_1\sigma) + 8bk^3 a_1\sigma = 0$$
$$\phi^3 : 2ca_2 + ak(2b_0 a_2 + b_1 a_1 + 2b_2 a_2\sigma) + 40bk^3 a_2\sigma = 0 \qquad (25)$$
$$\phi^4 : ak(2b_1 a_2 + b_2 a_1) + 6bk^3 a_1 = 0$$
$$\phi^5 : 2akb_2 a_2 + 24bk^3 a_2 = 0$$

For the second equation

$$\left.\begin{aligned}
&\phi^0 : cb_1\sigma + dka_0b_1\sigma + 2b_1k^3b\sigma^2 = 0 \\
&\phi^1 : 2cb_2\sigma + dk(2a_0b_2 + b_1a_1)\sigma + 16b_2bk^3\sigma^2 = 0 \\
&\phi^2 : cb_1 + dk(a_0b_1 + 2a_1b_2\sigma + a_2b_1) + 8bk^3b_1\sigma = 0 \\
&\phi^3 : 2cb_2 + dk(2a_0b_2 + b_1a_1 + 2b_2a_2\sigma) + 40bk^3b_2\sigma = 0 \\
&\phi^4 : dk(2a_1b_2 + a_2b_1) + 6bk^3b_1 = 0 \\
&\phi^5 : 2dkb_2a_2 + 24bk^3b_2 = 0
\end{aligned}\right\} \quad \ldots \quad (26)$$

Solving the above two system we get

$$a_0 = -\frac{c+8b\sigma k^3}{dk}, a_1 = 0, a_2 = -\frac{12bk^2}{d}$$

$$b_0 = -\frac{c+8b\sigma k^3}{ak}, b_1 = 0, b_2 = -\frac{12bk^2}{a}$$

Hence the general solution is

$$u(x,y,t) = \begin{cases} \left.\begin{aligned} a_0 - a_2\sigma \tanh^2\left(\sqrt{-\sigma}(ct+kx)\right) \\ a_0 - a_2\sigma \coth^2\left(\sqrt{-\sigma}(ct+kx)\right) \end{aligned}\right\} & \text{for} \quad \sigma < 0 \\ a_0 + a_2\left(\frac{(\Gamma(1+\alpha))^2}{((ct+kx)+\omega)^2}\right) & \text{for} \quad \sigma = 0, \omega - \text{Constant}.... \quad (27) \\ \left.\begin{aligned} a_0 + a_2\sigma \tan^2\left(\sqrt{\sigma}(ct+kx)\right) \\ a_0 + a_2\sigma \coth^2\left(\sqrt{\sigma}(ct+kx)\right) \end{aligned}\right\} & \text{for} \quad \sigma > 0 \end{cases}$$

$$v(x,y,t) = \begin{cases} \left.\begin{aligned} b_0 - b_2\sigma \tanh^2\left(\sqrt{-\sigma}(ct+kx)\right) \\ b_0 - b_2\sigma \coth^2\left(\sqrt{-\sigma}(ct+kx)\right) \end{aligned}\right\} & \text{for} \quad \sigma < 0 \\ b_0 + b_2\left(\frac{(\Gamma(1+\alpha))^2}{((ct+kx)+\omega)^2}\right) & \text{for} \quad \sigma = 0, \omega - \text{Constant}.... \quad (28) \\ \left.\begin{aligned} b_0 + b_2\sigma \tan_\alpha^2\left(\sqrt{\sigma}(ct+kx)\right) \\ b_0 + b_2\sigma \coth_\alpha^2\left(\sqrt{\sigma}(ct+kx)\right) \end{aligned}\right\} & \text{for} \quad \sigma > 0 \end{cases}$$

**6.0 Solutions of the Coupled fractional order KdV equations by fractional sub equation method**

Consider the coupled KdV equations with constant coefficients in the form

$$u_t^{(\alpha)} + avu_x^{(\alpha)} + bu_{xxx}^{(3\alpha)} = 0 \Big\} \tag{29}$$
$$v_t^{(\alpha)} + duv_x^{(\alpha)} + bv_{xxx}^{(3\alpha)} = 0 \Big\}$$

where $u_x^{(\alpha)} = \frac{\partial^\alpha u}{\partial x^\alpha}, u_t^{(\alpha)} = \frac{\partial^\alpha u}{\partial t^\alpha}, u_{xxx}^{(3\alpha)} = \frac{\partial^{3\alpha} u}{\partial x^{3\alpha}}$ and similar for $v$.

Again using one dimensional travelling wave transformation $\xi = kx + ct$ the equation (29) reduces to,

$$c^\alpha u_\xi^\alpha + ak^\alpha vu_\xi^\alpha + bk^{3\alpha} u_{\xi\xi\xi}^{3\alpha} = 0 \Big\} \tag{30}$$
$$c^\alpha v_\xi^\alpha + dk^\alpha uv_\xi^\alpha + bk^{3\alpha} v_{\xi\xi\xi}^{3\alpha} = 0 \Big\}$$

We want to find the series solution of the system of fractional differential equation in the following form where $\varphi(\xi)$ satisfies the fractional Riccati equation $D^\alpha \phi(\xi) = \sigma + \phi^2$, $0 < \alpha \leq 1$ and $a_i$'s are arbitrary constants [3].

$$u(\xi) = S_1(\varphi) = a_0 + a_1\varphi + a_2\varphi^2 + \ldots + a_n\varphi^n$$
$$v(\xi) = S_2(\varphi) = b_0 + b_1\varphi + b_2\varphi^2 + \ldots + b_n\varphi^n$$

Using the homogeneous balance principle as previous we get $m = n = 2$. Thus solution of (19) are of the form

$$u(\xi) = S_1(\varphi) = a_0 + a_1\varphi + a_2\varphi^2$$
$$v(\xi) = S_2(\varphi) = b_0 + b_1\varphi + b_2\varphi^2$$

Putting the above in equation (31) we get,

$$\left.\begin{array}{l} c^\alpha(a_1\sigma + 2a_2\sigma\varphi + a_1\varphi^2 + 2a_2\varphi^3) + ak^\alpha(b_0 + b_1\varphi + b_2\varphi^2)(a_1\sigma + 2a_2\sigma\varphi + a_1\varphi^2 + 2a_2\varphi^3) \\ +bk^{3\alpha}(2a_1\sigma^2 + 16a_2\varphi\sigma^2 + 8a_1\sigma\varphi^2 + 40a_2\sigma\varphi^3 + 6a_1\varphi^4 + 24a_2\varphi^5) = 0 \\ \\ \text{and } c^\alpha(b_1\sigma + 2b_2\sigma\varphi + b_1\varphi^2 + 2b_2\varphi^3) + ak^\alpha(b_0 + b_1\varphi + b_2\varphi^2)(b_1\sigma + 2b_2\sigma\varphi + b_1\varphi^2 + 2b_2\varphi^3) \\ +bk^{3\alpha}(2b_1\sigma^2 + 16b_2\varphi\sigma^2 + 8b_1\sigma\varphi^2 + 40b_2\sigma\varphi^3 + 6b_1\varphi^4 + 24b_2\varphi^5) = 0 \end{array}\right\} \ldots \tag{31}$$

Comparing the like powers of $\varphi$ we get

For the first equation

$$\begin{aligned}
\varphi^0 &: c^\alpha a_1 \sigma + ak^\alpha b_0 a_1 \sigma + 2a_1 k^{3\alpha} b\sigma^2 = 0 \\
\varphi^1 &: 2c^\alpha a_2 \sigma + ak^\alpha (2b_0 a_2 + b_1 a_1)\sigma + 16 a_2 bk^{3\alpha} \sigma^2 = 0 \\
\varphi^2 &: c^\alpha a_1 + ak^\alpha (b_0 a_1 + 2b_1 a_2 \sigma + b_2 a_1 \sigma) + 8bk^{3\alpha} a_1 \sigma = 0 \\
\varphi^3 &: 2c^\alpha a_2 + ak^\alpha (2b_0 a_2 + b_1 a_1 + 2b_2 a_2 \sigma) + 40 bk^{3\alpha} a_2 \sigma = 0 \\
\varphi^4 &: ak^\alpha (2b_1 a_2 + b_2 a_1) + 6bk^{3\alpha} a_1 = 0 \\
\varphi^5 &: 2ak^\alpha b_2 a_2 + 24 bk^{3\alpha} a_2 = 0
\end{aligned} \right\} \quad \ldots \quad (32)$$

For the second equation

$$\begin{aligned}
\varphi^0 &: c^\alpha b_1 \sigma + dk^\alpha a_0 b_1 \sigma + 2b_1 k^{3\alpha} b\sigma^2 = 0 \\
\varphi^1 &: 2c^\alpha b_2 \sigma + dk^\alpha (2a_0 b_2 + b_1 a_1)\sigma + 16 b_2 bk^{3\alpha} \sigma^2 = 0 \\
\varphi^2 &: c^\alpha b_1 + dk^\alpha (a_0 b_1 + 2a_1 b_2 \sigma + a_2 b_1) + 8bk^{3\alpha} b_1 \sigma = 0 \\
\varphi^3 &: 2c^\alpha b_2 + dk^\alpha (2a_0 b_2 + b_1 a_1 + 2b_2 a_2 \sigma) + 40 bk^{3\alpha} b_2 \sigma = 0 \\
\varphi^4 &: dk^\alpha (2a_1 b_2 + a_2 b_1) + 6bk^{3\alpha} b_1 = 0 \\
\varphi^5 &: 2dk^\alpha b_2 a_2 + 24 bk^{3\alpha} b_2 = 0
\end{aligned} \right\} \quad \ldots \quad (33)$$

Solving the above two system we get

$$a_0 = -\frac{c^\alpha + 8b\sigma k^{3\alpha}}{dk^\alpha}, a_1 = 0, a_2 = -\frac{12bk^{2\alpha}}{d}$$

$$b_0 = -\frac{c^\alpha + 8b\sigma k^{3\alpha}}{ak^\alpha}, b_1 = 0, b_2 = -\frac{12bk^{2\alpha}}{a}$$

Hence the general solution is

$$u(x,y,t) = \begin{cases} \begin{aligned} a_0 - a_2 \sigma \tanh_\alpha^2 \left(\sqrt{-\sigma}(c^\alpha t + k^\alpha x)\right) \\ a_0 - a_2 \sigma \coth_\alpha^2 \left(\sqrt{-\sigma}(c^\alpha t + k^\alpha x)\right) \end{aligned} & \text{for} \quad \sigma < 0 \\ a_0 + a_2 \left(\dfrac{(\Gamma(1+\alpha))^2}{((c^\alpha t + k^\alpha x)^\alpha + \omega)^2}\right) & \text{for} \quad \sigma = 0, \omega - \text{Constant} \\ \begin{aligned} a_0 + a_2 \sigma \tan_\alpha^2 \left(\sqrt{\sigma}(c^\alpha t + k^\alpha x)\right) \\ a_0 + a_2 \sigma \cot_\alpha^2 \left(\sqrt{\sigma}(c^\alpha t + k^\alpha x)\right) \end{aligned} & \text{for} \quad \sigma > 0 \end{cases} \quad \ldots \quad (34)$$

$$v(x,y,t) = \begin{cases} \begin{matrix} b_0 - b_2\sigma \tanh_\alpha^2\left(\sqrt{-\sigma}(c^\alpha t + k^\alpha x)\right) \\ b_0 - b_2\sigma \coth_\alpha^2\left(\sqrt{-\sigma}(c^\alpha t + k^\alpha x)\right) \end{matrix} & \text{for} \quad \sigma < 0 \\ b_0 + b_2\left(\dfrac{(\Gamma(1+\alpha))^2}{((c^\alpha t + k^\alpha x)^\alpha + \omega)^2}\right) & \text{for} \quad \sigma = 0, \omega - \text{Constant} \\ \begin{matrix} b_0 + b_2\sigma \tan_\alpha^2\left(\sqrt{\sigma}(c^\alpha t + k^\alpha x)\right) \\ b_0 + b_2\sigma \cot_\alpha^2\left(\sqrt{\sigma}(c^\alpha t + k^\alpha x)\right) \end{matrix} & \text{for} \quad \sigma > 0 \end{cases} \quad \dots (35)$$

## 7.0 Numerical simulation

In this section numerical simulations are done to find the solution pattern for different values of order of derivative $\alpha$. Here we are considering $c=1, k=1, b=a=1, d=1$. The numerical simulation is done for the solution set (20) and (34-35) for different values of order of derivative $\alpha$.

Since solution $u(x, y, t)$ in (20) is function of $x, y$ and $t$. The figures are drowning below for fixed value of $t = 1$ and different values of order of fractional derivative $\alpha$ for $\sigma < 0$.

Since the solutions $u$ and $v$ in (27, 28) and (34-35) become same under the considered values of the parameters. So here graphical presentation of $u$ only presented here in figure-2 for different values of order of fractional derivative $\alpha$ for $\sigma < 0$.

From the figure-1 and 2 it is clear that the shock solution for KP-Burger equation and soliton solution for coupled KdV equations occurs for $\alpha = 1$. With the decrease of $\alpha$ the solution patterns changes to periodic nature.

## 8.0 Conclusions

In this paper we found the analytical solutions of the non-linear partial differential equation integer order and Jumarie type fractional order partial differential equations. In Generalized tanh method the solutions of integer order non-linear partial differential equations are expressed in terms the hyperbolic functions (for $\sigma < 0$) and the trigonometric functions (for $\sigma > 0$) where as fractional sub-equation method express the solutions of non-linear partial differential equations in terms the fractional hyperbolic functions (for $\sigma < 0$) and the fractional trigonometric functions (for $\sigma > 0$). Both the methods are based on homogeneous balance principle. The solutions obtained in these methods are exact. From figure-1 it is clear that for small values of $\alpha$ in [0.65,1) there is a shock waves that include oscillation and when order of derivative tends to 1 the oscillation diminish. From figure-2 it is clear that the oscillatory solutions arises for small values of $\alpha$ in (<1) and that solution tends towards the soliton solution when order of derivative tends to 1. The solution obtained for $\sigma < 0$ matches with physical solutions of the KP-Burger equation and coupled KdV equations. As we know the Burger term is responsible for shock solution and due to the effect of dispersion in the medium.

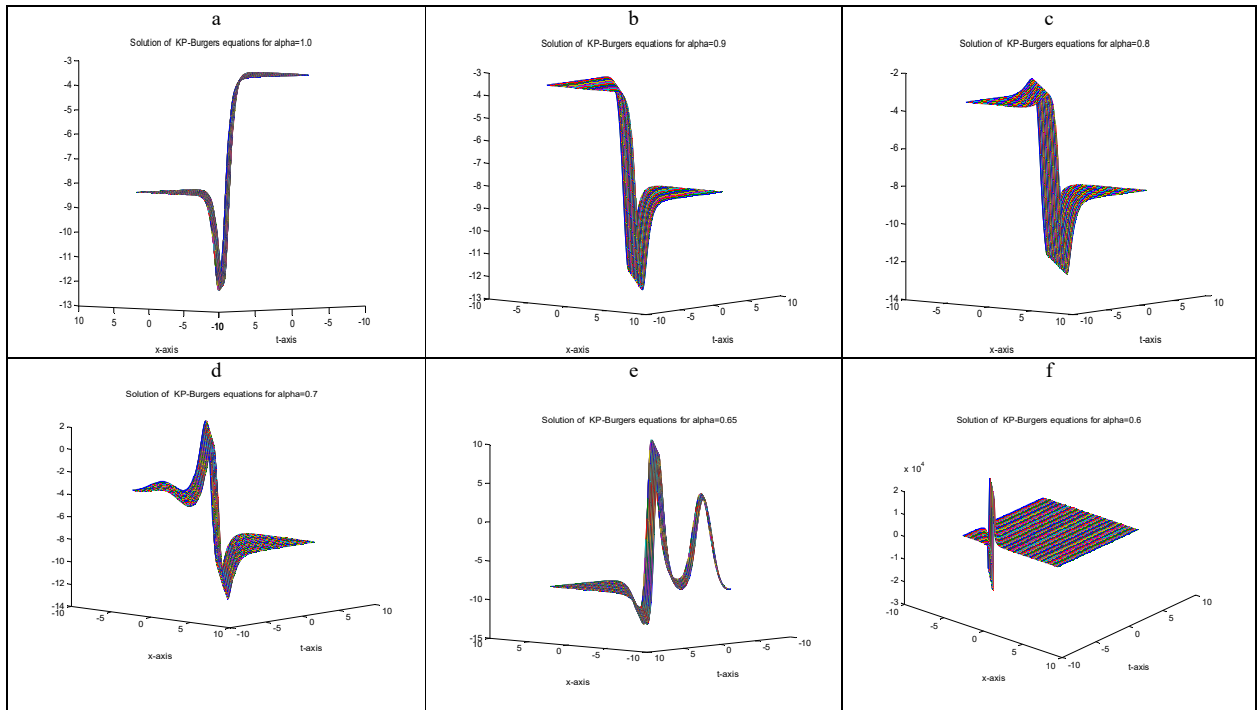

Fig-1: Graphical presentation of solutions for KP-Burger equation for $\sigma < 0$ for different values of order of fractional derivative $\alpha$.
(a) $\alpha = 1$ (b) $\alpha = 0.9$ (c) $\alpha = 0.8$ (d) $\alpha = 0.7$ (e) $\alpha = 0.65$ (f) $\alpha = 0.6$

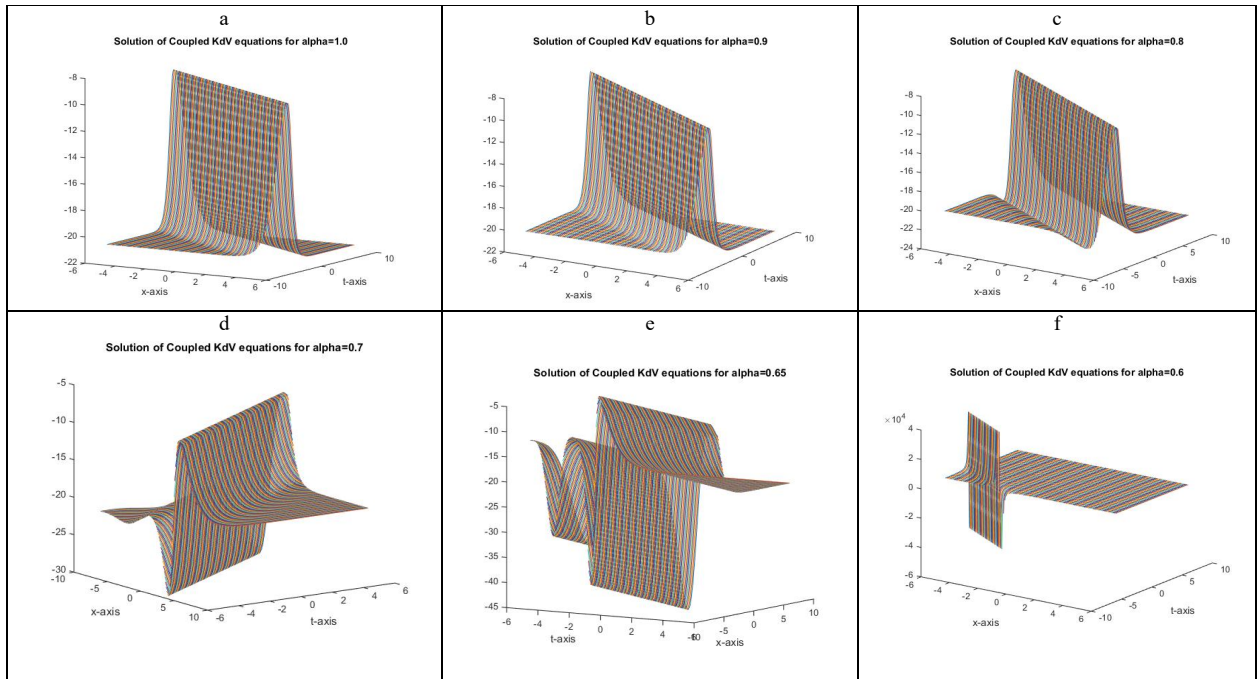

Fig-2: Graphical presentation of solutions for coupled KdV equations for $\sigma < 0$ for different values of order of fractional derivative $\alpha$.
(a) $\alpha = 1$ (b) $\alpha = 0.9$ (c) $\alpha = 0.8$ (d) $\alpha = 0.7$ (e) $\alpha = 0.65$ (f) $\alpha = 0.6$